\documentclass[prl,aps,superscriptaddress,unsortedaddress,twocolumn,%
  showpacs,preprintnumbers,amsmath,amssymb]{revtex4-1}
\usepackage{graphicx}
\usepackage[breaklinks,colorlinks=true]{hyperref}
\hypersetup{linkcolor=cyan, citecolor=blue}

\newcommand{\diag}{\mathrm{diag}}
\newcommand{\calT}{\mathcal{T}}
\newcommand{\calJ}{\mathcal{J}}
\newcommand{\calO}{\mathcal{O}}
\newcommand{\calS}{\mathcal{S}}
\newcommand{\Jcan}{J_{\text{can}}}

\newcommand{\bj}{\boldsymbol{j}}

\newcommand{\bv}{\boldsymbol{v}}
\newcommand{\bS}{\boldsymbol{S}}
\newcommand{\bnabla}{\boldsymbol{\nabla}}

\begin{document}
\title{Spin Hydrodynamics and Symmetric Energy-Momentum Tensors\\
-- A current induced by the spin vorticity --}
\author{Kenji Fukushima}
\affiliation{Department of Physics, The University of Tokyo,
  7-3-1 Hongo, Bunkyo-ku, Tokyo 113-0033, Japan}
\author{Shi Pu}
\affiliation{Department of Modern Physics, University of Science and
  Technology of China, Anhui 230026, China}

\begin{abstract}
  We discuss a puzzle in relativistic spin hydrodynamics; in the
  previous formulation the spin source from the antisymmetric part of
  the canonical energy-momentum tensor (EMT) is crucial.  The
  Belinfante improved EMT is pseudo-gauge transformed from the
  canonical EMT and is usually a physically sensible choice especially
  when gauge fields are coupled as in magnetohydrodynamics, but the
  Belinfante EMT has no antisymmetric part.  We find that
  pseudo-transformed entropy currents are physically inequivalent in
  nonequilibrium situations.  We also identify a current induced by
  the spin vorticity read from the Belinfante symmetric EMT{}.
\end{abstract}

\maketitle

\paragraph*{Introduction:}

Polarization measurements of $\Lambda$ and $\bar{\Lambda}$ baryons in
the relativistic heavy-ion collision have attracted lots of theoretical
interest~\cite{Abelev:2007zk}, which is driven by a recent
confirmation of the global polarization of $\Lambda$ and
$\bar{\Lambda}$ that signifies
``the most vortical fluid''~\cite{STAR:2017ckg}, as predicted in
Ref.~\cite{Liang:2004ph} and thermally quantified in
Ref.~\cite{Becattini:2016gvu}.  The underlying physics is intuitive:
non-central collisions provide hot and dense matter with the orbital
angular momentum as large as $\sim 10^6\hbar$ at the collision energy
$\sqrt{s_{NN}}=200\;\text{GeV}$ with the impact parameter
$b\sim 10\;\text{fm}$~\cite{Jiang:2016woz,Deng:2016gyh}.  Because of
the spin-orbital coupling in relativistic systems, a finite angular
momentum can be transported from the orbital angular momentum (OAM) to
the spin angular momentum (SAM).  We note that only the total angular
momentum (TAM) is a conserved quantity associated with rotational 
symmetry in relativistic theories but the spin degrees of freedom is
dissipative in a relativistic fluid.  Thus, only the particle intrinsic
spin affects the thermal abundance at the last stage, but the
relativistic spin hydrodynamics is needed for thorough understanding
of the spin evolution from the initial condition.   There are actually
some theoretical estimates based on parton cross
section~\cite{Liang:2004ph,Liang:2004xn}, and we should clarify a
missing bridge between the partonic and the thermal estimates.  For
some developments, see recent
reviews~\cite{Gao:2020vbh,Ryblewski:2020oir} and the references
therein.

For the spin hydrodynamics pioneering works are found in
Refs.~\cite{Montenegro:2017rbu,Montenegro:2017lvf} where the
Lagrangian effective field theory is adopted to approach the spin
polarized medium.  In Ref.~\cite{Florkowski:2017ruc} the spin
hydrodynamic equations are derived from the kinetic equations for
particles and antiparticles with spin 1/2.  More recently, the spin
hydrodynamic equations have been derived from the decomposition of the
energy-momentum tensor (EMT) and the entropy current analysis in
Ref.~\cite{Hattori:2019lfp}.  See Ref.~\cite{Speranza:2020ilk} for
another approach and discussions in gravitational physics and also
Refs.~\cite{Singh:2020efc,Singh:2020yjn,Singh:2020zdn} for recent
discussions based on the Bjorken hydrodynamics.  Another possible hint
to build the spin hydrodynamics comes from the quantum kinetic theory
for massive fermions with collisions, e.g., see
Refs.~\cite{Gao:2019zhk, Weickgenannt:2019dks,Wang:2019moi,Liu:2020flb,Li:2019qkf,Weickgenannt:2020aaf,Yang:2020hri}.

In the present work we focus on a controversy in regard to
pseudo-gauge ambiguity of the EMT for the spin physics.  Similar
problems have been well known also in the context of the proton spin
decomposition;  conventionally, the spin decomposition based on the
canonical EMT is called the Jaffe-Manohar decomposition, while an
improved EMT that is symmetric and gauge invariant gives the Ji
decomposition (see a recent essay~\cite{Fukushima:2020qta} and
references therein).  In gauge theories such as quantum chromodynamics
(QCD) the symmetric EMT is manifestly gauge invariant, and is directly
related to physical observables.  Also in the future electron-ion
collider (EIC) experiment the proton EMT will be
measured~\cite{Burkert:2018bqq,Shanahan:2018nnv}.  For this the
symmetric EMT is empirically assumed.

The central puzzle in the context of the spin hydrodynamics lies in
the fact that the spin degrees of freedom seem to appear from the
antisymmetric component of the canonical EMT{}.  As we mentioned
above, however, the canonical EMT is not gauge invariant if gauge
fields are involved as in magnetohydrodynamics.  One might think of a
way to enforce gauge invariance on the canonical
EMT~\cite{Chen:2008ag,Wakamatsu:2010qj,Hatta:2011zs}, but such a
prescription requires nonlocal gauge potentials, which is not
systematically implemented in hydrodynamics.  Therefore, it would be
theoretically preferable to formulate the spin hydrodynamics based on
the symmetric EMT or the Belinfante improved form of the EMT after an
appropriate pseudo-gauge transformation.  The technical problem is,
however, that one can no longer extract the spin part from the
antisymmetric component that identically vanishes in the Belinfante
form.  It is a common consensus that the canonical and the Belinfante
forms are both qualified as physical EMTs, and yet only the canonical
EMT works for the derivation of the spin hydrodynamics as employed in
Ref.~\cite{Hattori:2019lfp}.  We note that some inequivalent
properties between different EMTs have been revealed in
nonequilibrium environments~\cite{Becattini:2011ev,Becattini:2012pp, Becattini:2014yxa},
but this difference would not necessarily exclude a possibility to
derive the spin hydrodynamics from the Belinfante EMT{}.  Actually we
will pursue this possibility and eventually reach a conclusion to
support discussions by those preceding works in an illuminating way.

Here, let us summarize our notation.  The metric is
$g_{\mu\nu}=\diag(+,-,-,-)$ and the projection operator in our
convention is
$\Delta^{\mu\nu}\equiv g^{\mu\nu}-u^\mu u^\nu$ with the four-vector
fluid velocity $u^\mu$.  We use $\calT^{\mu\nu}$ to represent the
Belinfante EMT, while $\Theta^{\mu\nu}$ is the canonical one.  Also,
we define $\calJ^{\mu\alpha\beta}$ for the TAM in the Belinfante form
and $\Jcan^{\mu\alpha\beta}$ for the TAM in the canonical form.  For
an arbitrary tensor $A^{\mu\nu}$, we define its symmetric and
antisymmetric parts as
$A_{\rm (s)}^{\mu\nu}= A^{(\mu,\nu)}
\equiv \tfrac{1}{2}(A^{\mu\nu}+ A^{\nu\mu})$ and
$A_{\rm (a)}^{\mu\nu}= A^{[\mu, \nu]}
\equiv \tfrac{1}{2}(A^{\mu\nu}-A^{\nu \mu})$.  We also use a symbol,
$<...>$, to mean the traceless part, i.e., 
$A^{<\mu\nu>}\equiv 
\frac{1}{2}[\Delta^{\mu\alpha}\Delta^{\nu\beta}
+\Delta^{\nu\alpha}\Delta^{\mu\beta}]A_{\alpha\beta}
-\frac{1}{3}\Delta^{\mu\nu}(A^{\rho\sigma}\Delta_{\rho\sigma})$. 
\vspace{0.5em}

\paragraph*{Canonical vs.\ Belinfante formulations:}
To make our point clear we shall make a brief review of the spin
hydrodynamics from the canonical EMT as discussed in
Ref.~\cite{Hattori:2019lfp}.  In the canonical form the TAM can be
decomposed into
\begin{equation}
  \Jcan^{\alpha\mu\nu}=x^\mu \Theta^{\alpha\nu}-x^\nu\Theta^{\alpha\mu}
  +\Sigma^{\alpha\mu\nu}\,,
  \label{eq:can_decom}
\end{equation}
where $x^\mu\Theta^{\alpha\nu}-x^\nu\Theta^{\alpha\mu}$ and
$\Sigma^{\alpha\mu\nu}$ represent the OAM and the SAM tensors,
respectively.  From the conservation laws of the TAM and the EMT we
readily find,
\begin{equation}
  \partial_\alpha \Sigma^{\alpha\mu\nu} =
  -2\Theta_{\rm (a)}^{\mu\nu}
  \label{eq:spinEMT}
\end{equation}
with 
$\Theta_{\rm (a)}^{\mu\nu}$ being the antisymmetric part of the
canonical EMT, which is understood as spin nonconservation in
relativistic systems.

Recalling that the spin in the quantum field theory is
$\Sigma^{0ij}\sim S^{ij}=\epsilon^{ijk}S^k$, we can decompose the spin
tensor in terms of hydrodynamical variables as follows:
\begin{equation}
  \Sigma^{\alpha\mu\nu} = u^\alpha S^{\mu\nu}
  + \Sigma_{(1)}^{\alpha\mu\nu}\,.
  \label{eq:spin_decom}
\end{equation}
We can understand Eq.~\eqref{eq:spin_decom} in analogy to
decomposition of the charge current; $j^\mu =nu^\mu +j^\mu_{(1)}$
(where $u\cdot j_{(1)}=0$), with the charge density $n$ and the
dissipative current $j^\mu_{(1)}$ from the higher order in the
gradient expansion.  Correspondingly, we can identify $S^{\mu\nu}$ as
the spin density and $\Sigma_{(1)}^{\alpha\mu\nu}$ as the dissipative
higher order correction.  We can neglect $\Sigma_{(1)}^{\alpha\mu\nu}$
since only
$\partial_\alpha\Sigma_{(1)}^{\alpha\mu\nu} \sim \calO(\partial^2)$
appears that is beyond the order focused in this work.

In deriving the hydrodynamic equations the entropy current and the
second law of thermodynamics are essential.  For this purpose we need
to express the entropy current involving the spin tensors.  The
thermodynamic relation in equilibrium reads,
\begin{equation}
  e + p = Ts +\mu n+ \omega_{\mu\nu} S^{\mu\nu}\,,
  \label{eq:thermorelation}
\end{equation}
where $e$, $p$, $T$, $s$, and $\mu$ are the energy density, the
pressure, the temperature, the entropy density, and the chemical
potential, respectively.  We also introduced the spin potential,
$\omega_{\mu\nu}$, according to the prescription of
Ref.~\cite{Hattori:2019lfp}.  For simplicity, we only consider one
$U(1)$ conserved charge, e.g., the electric charge or the baryon
charge.  If necessary, we can easily extend our discussion to multiple
conserved charges.  For actual calculations differential forms of
Eq.~\eqref{eq:thermorelation} are convenient; namely,
$de=Tds+ \mu dn+\omega_{\mu\nu}d S^{\mu\nu}$ and 
$dp=sdT+n d\mu+ S^{\mu\nu}d \omega_{\mu\nu}$.  In the present
convention $e$ is a function of $S^{\mu\nu}$, while $p$ is a function
of $\omega^{\mu\nu}$.

Now, let us introduce a nonequilibrium entropy current
$\calS_{\text{can}}^\mu$ following a prescription of
Ref.~\cite{Israel:1979wp}.  In the presence of the spin density and
the spin potential we can postulate: 
\begin{align}
  \calS^\mu_{\text{can}}
  &= \frac{u_{\nu}}{T}\Theta^{\mu\nu}
  + \frac{p}{T}u^{\mu}-\frac{\mu}{T}j^{\mu}
  -\frac{1}{T}\omega_{\rho\sigma}S^{\rho\sigma}u^{\mu}
    +\calO(\partial^2) \notag\\
  &= s u^\mu + \frac{u_\nu}{T}\Theta_{(1)}^{\mu\nu}
    -\frac{\mu}{T} j_{(1)}^\mu + \calO(\partial^2)\,,
\label{eq:Scan}
\end{align}
where $\Theta_{(1)}^{\mu\nu}$ as well as $j_{(1)}^\mu$ denotes
dissipative terms.  This explicit form clearly shows that
the entropy current has a definite relation to the equilibrium entropy
up to the leading order, but the higher orders are not uniquely
constrained.  Therefore, Eq.~\eqref{eq:Scan} should be regarded as an
Ansatz.

Using Eq.~\eqref{eq:thermorelation} and
$u_\nu\partial_\mu\Theta^{\mu\nu}=0$, we can prove
$T\partial_\mu(su^\mu)-\mu\partial_\mu j_{(1)}^\mu
+\omega_{\rho\sigma}\partial_\mu(S^{\rho\sigma}u^\mu)
+u_\nu\partial_\mu\Theta_{(1)}^{\mu\nu}=0$.  This significantly
simplifies the divergence of the entropy current into
\begin{equation}
  \partial_\mu \calS_{\text{can}}^\mu 
  = -j_{(1)}^\mu \partial_\mu\frac{\mu}{T}
  -\frac{\omega_{\rho\sigma}}{T}\partial_{\mu}(u^{\mu}S^{\rho\sigma})
  +\Theta_{(1)}^{\mu\nu}\partial_{\mu}\frac{u_{\nu}}{T} \,.
  \label{eq:entropy_can}
\end{equation}
In the right-hand side we can use
$\partial_{\mu}(u^\mu S^{\rho \sigma})
= -2\Theta_{\rm (a)}^{\rho\sigma} + \calO(\partial^2)$ which comes from
Eqs.~\eqref{eq:spinEMT} and \eqref{eq:spin_decom}.  At the first
order, moreover, the tensor decomposition leads to
$\Theta_{(1)}^{\mu\nu}=\Theta_{\rm (1s)}^{\mu\nu}+\Theta_{\rm (1a)}^{\mu\nu}$
with
\begin{equation}
  \Theta_{\rm (1s)}^{\mu\nu} = 2 h^{(\mu} u^{\nu)} + \pi^{\mu\nu}\,,
  \qquad
  \Theta_{\rm (1a)}^{\mu\nu} = 2 q^{[\mu} u^{\nu]} + \phi^{\mu\nu}\,.
  \label{eq:theta1}
\end{equation}
As usual $\pi^{\mu\nu}$ is the viscous tensor and $\phi^{\mu\nu}$ is
its antisymmetric counterpart.  Likewise, $h^\mu$ is the heat flow and
$q^\mu$ is its counterpart in the antisymmetric sector.  In
calculational steps
$u_\mu\pi^{\mu\nu}=u_\mu\phi^{\mu\nu}=u\cdot q=u\cdot h=0$ will be
useful.  As discussed in Ref.~\cite{Hattori:2019lfp} we can collect
terms involving $\pi^{\mu\nu}$, $\phi^{\mu\nu}$, $h^\mu$, and $q^\mu$
and identify their tensorial forms from the sufficient condition for
the second law of thermodynamics,
$\partial_\mu\calS^\mu_{\text{can}}\ge 0$, as realized in a form of
sum of squares.

Then, $\pi^{\mu\nu}$ and $h^\mu$ are found to have no
spin corrections, while $q^\mu$ and $\phi^{\mu\nu}$ are found to have
terms $\propto \omega^{\mu\nu}$ as
\begin{align}
  q^\mu
    &= \lambda \bigl[ T^{-1}\Delta^{\mu\alpha}\partial_\alpha T
      + (u\cdot\partial) u^\mu - 4\omega^{\mu\nu} u_\nu \bigr]\,,
      \label{eq:q} \\
  \phi^{\mu\nu}
    &= -\gamma(\Omega^{\mu\nu} -
      2T^{-1}\Delta^{\mu\alpha}\Delta^{\nu\beta} \omega_{\alpha\beta})\,,
      \label{eq:phi}
\end{align}
where $\Omega^{\mu\nu}\equiv - \Delta^{\mu\rho}\Delta^{\nu\sigma}
\partial_{[\rho} (\beta u_{\sigma]})$ is usually referred to as the
thermal vorticity~\cite{Becattini:2014yxa}, and $\lambda$ and $\gamma$
are nonnegative transport coefficients.  We can reasonably understand
the physical interpretation:  The rotation carried by the fluid
velocity and the thermal gradient together with the spin chemical
potential plays a role of the source to produce/absorb the spin.
Then, the spin hydrodynamics dictates the evolution of
$\omega_{\mu\nu}$ or $S^{\mu\nu}$ and the local thermal equilibrium 
relation, $S^{\mu\nu}=\partial p/\partial\omega_{\mu\nu}|_{T,\mu}$,
imposes a connection between them.

From above discussions it is clear that Eq.~\eqref{eq:spinEMT} is
crucial for constructing hydrodynamics with spin degrees of freedom,
and it seems to be indispensable to keep $\Theta_{\rm (a)}^{\mu\nu}$.
The EMT, however, has pseudo-gauge invariance, and one can always
choose a symmetrized or Belinfante improved EMT form without losing
physics contents.
\vspace{0.5em}

\paragraph*{Spin strikes back:}
The confusion lies in the absence of the antisymmetric part of the
Belinfante EMT which implies no spin degrees of freedom at all.  We
obtain the symmetric Belinfante EMT by the following pseudo-gauge
transformation:
\begin{align}
  \calT^{\mu\nu} &= \Theta^{\mu\nu} + \partial_\lambda K^{\lambda\mu\nu}\,,
  \label{eq:bel_can}\\
  K^{\lambda\mu\nu} &= \frac{1}{2} \bigl( \Sigma^{\lambda\mu\nu}
  - \Sigma^{\mu\lambda\nu} + \Sigma^{\nu\mu\lambda} \bigr)\,. 
  \label{eq:K}
\end{align}
With this choice we can get rid of the spin source and it is easy to
confirm that $\calT^{\mu\nu}$ is symmetric;
$\calT^{\mu\nu}=\calT^{\nu\mu}$.  Here, $K^{\lambda\mu\nu}$ is
antisymmetric with respect to $\lambda$ and $\mu$, so that
$\partial_\mu \calT^{\mu\nu}=0$ still holds as long as $\partial_\mu
\Theta^{\mu\nu}=0$.  In other words we have an identity,
\begin{equation}
  \partial_\mu\partial_\lambda
  \bigl( u^\lambda S^{\mu\nu} + u^\mu S^{\nu\lambda}
  + u^\nu S^{\mu\lambda} \bigr) = 0\,,
  \label{eq:identityS}
\end{equation}
from Eqs.~\eqref{eq:spin_decom} and \eqref{eq:K}.  This equation
corresponds to the ``quantum spin vorticity principle'' in the quantum
spin vorticity theory~\cite{spinvorticity}.

The Belinfante improved TAM, which is a counterpart of
Eq.~\eqref{eq:can_decom}, reads,
\begin{equation}
  \calJ^{\alpha\mu\nu} = x^\mu \calT^{\alpha\nu} - x^\nu 
  \calT^{\alpha\mu}\,,
  \label{eq:bel_J}
\end{equation}
where $\calJ^{\alpha\mu\nu} \equiv J^{\alpha\mu\nu} + \partial_\rho
(x^\mu K^{\rho\alpha\nu} - x^\nu K^{\rho\alpha\mu})$.
Equation~\eqref{eq:bel_J} looks like an OAM relation [see the first
part in Eq.~\eqref{eq:can_decom}] and it is often said that the spin is
identically vanishing in the Belinfante form.  Precisely speaking,
since the energy-momentum conservation,
$\partial_{\mu} \calT^{\mu\nu}=0$, leads to the TAM conservation,
$\partial_{\alpha} \calJ^{\alpha\mu\nu}=0$, in the Belinfante form,
one cannot find a counterpart of Eq.~\eqref{eq:spinEMT}.  Our point is
that we do not have to go through the EMT to write down such a tensor
decomposition.

Before addressing the entropy analysis, we shall discuss a possibility
to introduce terms with $S^{\mu\nu}$ in the symmetric EMT form;  the
tensor indices we can use are not only $g^{\mu\nu}$, $u^\mu$,
$\partial^\mu$, but also $S^{\mu\nu}$ in general.  The guiding
principle is provided from a transformation between $\calT^{\mu\nu}$
and $\Theta^{\mu\nu}$.  We can utilize Eq.~\eqref{eq:bel_can} together
with $\Sigma^{\mu\alpha\beta}=u^\mu S^{\alpha\beta} + \calO(\partial)$, to
find,
\begin{align}
  \calT^{\mu\nu}
  &= \Theta^{\mu\nu} + \frac{1}{2}\partial_\lambda
    (u^\lambda S^{\mu\nu} - u^\mu S^{\lambda\nu}
    + u^\nu S^{\mu\lambda}) + \calO(\partial^2) \notag\\
  &= \Theta_{\rm (s)}^{\mu\nu} + \frac{1}{2} \bigl[ \partial_\lambda
    (u^\mu S^{\nu\lambda} + u^\nu S^{\mu\lambda})\bigr]
    + \calO(\partial^2)\,.
\end{align}
If we need to construct the hydrodynamics using the symmetric EMT as
demanded in the case with gauge fields, we must employ the above form
of symmetric spin corrections.  One might think 
that $\partial_\mu \calT^{\mu\nu}=0$ may look different from
$\partial_\mu \Theta^{\mu\nu}=0$, but they are equivalent thanks to
Eq.~\eqref{eq:identityS};  therefore, Eq.~\eqref{eq:identityS}
constitutes an evolution equation.  The hydrodynamic expansion leads
to
\begin{equation}
  \calT^{\mu\nu} = (e+p)u^\mu u^\nu - pg^{\mu\nu}
  + \calT_{(1)}^{\mu\nu} + \calO(\partial^2)\,,\label{eq:EMT_Bel}
\end{equation}
where
\begin{equation}
  \calT_{(1)}^{\mu\nu} = 2h^{(\mu} u^{\nu)} + \pi^{\mu\nu}
  + \frac{1}{2} \partial_\lambda (u^\mu S^{\nu\lambda}
  + u^\nu S^{\mu\lambda})\,. 
\end{equation}
We should emphasize that $\calT_{(1)}^{\mu\nu}$ is still symmetric
with respect to $\mu$ and $\nu$ even with spin involving terms.

The heat flow, $h^\mu$, is defined from the symmetric index structure
involving $u^\nu$.  Therefore, once $\calT_{(1)}^{\mu\nu}$ is
given, one can identify $h^\mu$ from the tensor decomposition of
$\calT_{(1)}^{\mu\nu}$.  In the presence of spin correction terms, the
tensor decomposition leads to the heat flow coupled to the spin.  We
can readily see this from the following decomposition:
\begin{equation}
  \begin{split}
    &2h^{(\mu} u^{\nu)}  + \pi^{\mu\nu} + \frac{1}{2} \bigl[ \partial_\lambda 
    (u^\mu S^{\nu\lambda} + u^\nu S^{\mu\lambda}) \bigr] \\
    & = \delta e u^\mu u^\nu 
    + 2\bigl( h^{(\mu} + \delta h^{(\mu}\bigr) u^{\nu)}
    + \pi^{\mu\nu} + \delta \pi^{\mu\nu}\,.
  \end{split}
  \label{eq:spin_corr}
\end{equation}
Here, we have the energy density correction, $\delta e$, the heat flow
correction, $\delta h^\mu$, and the viscous tensor correction,
$\delta\pi^{\mu\nu}$, given respectively by
\begin{equation}
  \begin{split}
  \delta e
  & =  u_{\rho}\partial_{\sigma}S^{\rho\sigma}\,,\\
  \delta h^{\mu}
  & = \frac{1}{2} \bigl[\Delta_{\sigma}^{\mu}
    \partial_{\lambda}S^{\sigma\lambda}
    +u_{\rho}S^{\rho\lambda}\partial_{\lambda}u^{\mu}\bigr]\,,\\
  \delta\pi^{\mu\nu}
  & = \partial_{\lambda}(u^{<\mu}S^{\nu>\lambda})
    + \delta\Pi \Delta^{\mu\nu}\,,\\
  \delta\Pi
  & = \frac{1}{3}\partial_{\lambda}
    (u^{\sigma}S^{\rho\lambda})\Delta_{\rho\sigma}\,,
  \end{split}
  \label{eq:spin_corr}
\end{equation}
where $\delta\Pi$ is the bulk viscous correction.  We note that the
above correction of $\delta h^\mu$ is consistent with the momentum
density induced by the spin vorticity that has been discussed in the
quantum spin vorticity theory~\cite{spinvorticity}.  We will later
discuss the physical meaning in more details.
  
One may wonder how $q^\mu$ and $\phi^{\mu\nu}$ can be retrieved in the
Belinfante formalism at all, since they are extracted from the antisymmetric
EMT as in Eq.~\eqref{eq:theta1}, which is identically vanishing in the
Belinfante form.  As we exercised for the canonical EMT, let us
consider the entropy current.  The Belinfante counterpart of the
thermodynamic extension~\eqref{eq:Scan} reads,
\begin{align}
  \calS^{\mu}
  &=  \frac{u_{\nu}}{T}\calT^{\mu\nu}+\frac{p}{T}u^{\mu}
    -\frac{\mu}{T}j^{\mu}-\frac{1}{T}\omega_{\rho\sigma}
    S^{\rho\sigma}u^{\mu}+\mathcal{O}(\partial^{2}) \notag\\
  & =  su^{\mu} + \frac{u_\nu}{T}\calT_{(1)}^{\mu\nu}
    - \frac{\mu}{T} j_{(1)}^{\mu} + \calO(\partial^2)
    \label{eq:SBel}
\end{align}
with which the divergence of the entropy current takes the following
form:
\begin{equation}
  \partial_{\mu}\calS^{\mu}
  =  \biggl( \frac{n}{e+p} h^\mu - j_{(1)}^{\mu}\biggr)  
  \Delta_{\mu\nu}\partial^{\nu}\frac{\mu}{T}
  +\frac{1}{T}\pi^{\mu\nu}\partial_{\mu}u_{\nu} + \Delta
  \label{eq:entropy_Bel1}
\end{equation}
with
\begin{equation}
  \Delta \equiv
  \frac{1}{2} \bigl[\partial_{\lambda}(u^{\mu}S^{\nu\lambda}+u^{\nu}S^{\mu\lambda}) \bigr]
  \partial_{\mu}\frac{u_{\nu}}{T} - \frac{\omega_{\rho\sigma}}{T}
  \partial_\lambda (u^\lambda S^{\rho\sigma})\,.  
  \label{eq:entropy_Bel2}
\end{equation} 
Here, we emphasize that Eqs.~(\ref{eq:entropy_Bel1},
\ref{eq:entropy_Bel2}) are \textit{not} equivalent to
Eq.~\eqref{eq:entropy_can} even with Eq.~\eqref{eq:identityS}.  For
more clarification we will transform Eq.~\eqref{eq:entropy_Bel2}
using Eq.~\eqref{eq:identityS}.  We can add Eq.~\eqref{eq:identityS} to find,
\begin{equation}
  \begin{split}
    \Delta &= \frac{1}{2}\partial_\mu \biggl[ \partial_\lambda
    (u^\lambda S^{\mu\nu} + u^\mu S^{\nu\lambda} + u^\nu S^{\mu\lambda}) \frac{u_\nu}{T}
    \biggr] \\
    &\quad\qquad
    - \frac{1}{2} \bigl[\partial_\lambda (u^\lambda S^{\mu\nu})\bigr]
    \partial_\mu \frac{u_\nu}{T} - \frac{\omega_{\rho\sigma}}{T}
    \partial_\lambda (u^\lambda S^{\rho\sigma})\,. 
  \end{split}
\end{equation}
Therefore, the difference between
Eqs.~(\ref{eq:entropy_Bel1}, \ref{eq:entropy_Bel2}) and
\eqref{eq:entropy_can} turns out to be a total derivative.  We recall
that Eq.~\eqref{eq:SBel} is an Ansatz and we could have defined an
entropy, $\calS'$, to absorb the total derivative and then we arrive
at
\begin{equation}
  \Delta' =
  -\partial_{\lambda}(u^{\lambda}S^{\mu\nu})\biggl(
  \frac{1}{2}\partial_{\mu}\frac{u_{\nu}}{T}+\frac{\omega_{\mu\nu}}{T}\biggr)\,.
\end{equation}
In this case $\partial_\mu \calS^{\prime\mu}$ is given by
Eq.~\eqref{eq:entropy_Bel1} with $\Delta$ replaced by $\Delta'$.
Interestingly, $\calS^{\prime\mu}$ is just the same as
$\calS_{\text{can}}^\mu$.  Then, the constrains from the entropy
principle amount to those in the canonical formulation from
Eqs.~\eqref{eq:theta1}, \eqref{eq:q}, and \eqref{eq:phi}.  In
principle, alternatively, one may constrain $S^{\mu\nu}$ directly
from Eq.~\eqref{eq:entropy_Bel2} employing the following tensor
decomposition:
\begin{equation}
  S^{\mu\nu} = 2 \mathfrak{s}^{[\mu} u^{\nu]}
  - \epsilon^{\mu\nu\rho\sigma}u_\rho S_\sigma 
  \label{eq:Sdecom}
\end{equation}
with $u\cdot \mathfrak{s}=0$.  We have tried but this is a difficult 
task to constrain $\mathfrak{s}^\mu$ and $S^\mu$ from the 
entropy principle due to the presence of derivatives.  The difficulty 
seems to favor the canonical choice of $\calS^\mu$. 

We emphasize that such a difference by the total derivative is
irrelevant to bulk thermodynamics properties and a stringent condition
of the local thermal equilibrium gives rise to the physical difference
in the entropy current.  We make a remark here;  this total derivative
shift is quite analogous to $V^\mu$ in
Refs.~\cite{Glorioso:2016gsa,Glorioso:2017fpd}.  There, the shift by
$V^\mu$ appears from the dynamical KMS condition in the effective
field theory approach to hydrodynamics.  It would be a very
interesting future work to pursue a possible relationship.  Our
transformation from $\calS^\mu$ to $\calS^{\prime\mu}$ is actually
analogous to the hydrodynamical treatment of the triangle anomaly in
Ref.~\cite{Son:2009tf}, where the EMT is also symmetric and some terms
proportional to the vorticity are added to the entropy flow.  We also
emphasize that our observation is consistent with the claim made in
Refs.~\cite{Becattini:2011ev,Becattini:2012pp}.  They found using the
density operator that the canonical and the Belinfante EMTs are
equivalent only in equilibrium but they are \textit{not} in
nonequilibrium systems~\cite{Becattini:2018duy}.  In our analysis the
pseudo-gauge transformation generates conserved EMTs and leads to
different expressions for the entropy current.  With those different
expressions the physics is not equivalent once we take account of
dissipative terms and impose the second law of thermodynamics,
$\partial_\mu\calS^\mu\ge 0$, for dynamics out of equilibrium.
\vspace{0.5em}

\paragraph*{Physical interpretation of spin correction terms:}

We have seen that we must introduce a modified entropy current and
then the entropy principle supports the canonical results in
Eqs.~\eqref{eq:q} and \eqref{eq:phi}.  Nevertheless, we emphasize that
the Belinfante EMT should be physical and the spin corrections by
Eq.~\eqref{eq:spin_corr} are physical as well.  We must be, however,
careful of the physical interpretation in relativistic hydrodynamics.
The heat flow correction by $\delta h^\mu$, for example, is not
physical by itself.

In relativistic hydrodynamics $u^\mu$ is not unique in general and
one should make a choice of the frame;  the common choice is the
Landau frame or the energy frame.  Then, in this frame, the heat flow
is absent by construction.  More specifically, we should impose the
Landau condition for the relativistic hydrodynamics and choose the
fluid velocity $u_{\rm L}^\mu$ to satisfy
 $\Delta_{\rho\mu}^{\rm L}\calT^{\mu\nu}_{\rm L} u_{{\rm L}\nu}=0$,
where ``L'' denotes the quantities in the Landau frame.  We can
introduce the fluid velocity, $u_{\rm L}^\mu$, as
\begin{equation}
  u_{\rm L}^\mu = u^\mu + \frac{1}{e+p}
  (h^\mu + \delta h^\mu)\,.
  \label{eq:Landau_frame}
\end{equation}
We can also transfer the Belinfante EMT in Eq.~\eqref{eq:EMT_Bel} to
the one in the Landau frame as
$\calT_{\rm L}^{\mu\nu} = (e+\delta e)u_{\rm L}^\mu u_{\rm L}^\nu
-(p+\delta\Pi)\Delta_{\rm L}^{\mu\nu}+\pi_{\rm L}^{\mu\nu}
+\delta\pi_{\rm L}^{\mu\nu}+\calO(\partial^2)$ and there is no term
corresponding to the heat flow.

In this frame with the fluid velocity given by
Eq.~\eqref{eq:Landau_frame} the heat flow is absent but the modified
current remains finite, which reads:
\begin{equation}
  j^\mu_{{\rm L}(1)} = \biggl( j^\mu_{(1)} - \frac{n}{e+p}h^\mu\biggr)
  + \delta j_{(1)}^\mu
\end{equation}
with
\begin{equation}
  \delta j_{(1)}^\mu = -\frac{n}{e+p} \delta h^\mu \,.
\end{equation}
The first part in the parentheses,
$(j^\mu_{(1)} - \frac{n}{e+p}h^\mu)$, is an invariant combination in
different frames~\cite{Israel:1979wp}, which also appeared in
Eq.~\eqref{eq:entropy_Bel1}.  We can represent the induced current in
terms of the spin or the decomposed form in Eq.~\eqref{eq:Sdecom}.
Namely, Eq.~\eqref{eq:Sdecom} gives $\mathfrak{s}^i=S^{i0}$ and
$S^i=\frac{1}{2}\epsilon^{ijk}S^{jk}$.  The complete expression of
$\delta h^\mu$ in Eq.~\eqref{eq:spin_corr} involves many terms, and we
can simplify them by taking the nonrelativistic reduction of
$u^\mu=(1,\bv)$ with $\bv\to 0$, while derivatives of $\bv$ are still
kept nonvanishing.  Then, in the three-vector representation, we find,
\begin{equation}
  \begin{split}
    \delta\bj_{(1)}
    = -\frac{n}{2(e+p)}\bigl[ &\bnabla\times\bS + \dot{\bv}\times\bS\\
    &+ (\bnabla\cdot\bv)\mathfrak{s}
    - 2(\mathfrak{s}\cdot\bnabla)\bv + \dot{\mathfrak{s}}\bigr]\,.
  \end{split}
\end{equation}
One may think that the overall sign is opposite to that in the quantum
spin vorticity theory~\cite{spinvorticity}.  This difference is
attributed to the frame choice.  We are working in a frame comoving
with the heat flow, and this reverses the overall direction of the
induced current.
\vspace{0.5em}

\paragraph*{Summary:}
We formulated the spin hydrodynamics using the symmetric EMT which is
commonly considered to be physical.  The added terms satisfy an
identity for the spin tensor which corresponds to the quantum spin
vorticity principle.  The equations of motion are equivalent, but we
found that the entropy analysis makes an inequivalent deviation.  The
entropy current derived from the canonical formulation is different
from the one from the symmetric EMT by a total derivative.  Therefore,
if we impose a constraint not globally but \textit{locally} from the
second law of thermodynamics, the pseudo-gauge transformation would
lead to different physical contents in nonequivalent systems.  With
our formulation based on the symmetric EMT, we established a relation
between the spin vorticity (i.e., the rotation of the spin) and the
(electric) current, $\delta\bj\propto \bnabla\times\bS$, in a
hydrodynamical way.  One may find a similar
relation using the Dirac equation in quantum field theories, and our
formula is more complete with fluid velocity terms.  Applications to
the heavy-ion phenomenology should deserve further investigations.
\vspace{1em}

\begin{acknowledgments}
  The authors thank
  Francesco~Becattini,
  Wojciech~Florkowski, and
  Xu-Guang~Huang
  for stimulating discussions at a workshop,
  ``New Development of Hydrodynamics and its Applications in Heavy-Ion
  Collisions,'' on Oct.~30-Nov.~2, 2019.
  K.~F.\ was supported by Japan Society for the Promotion
  of Science (JSPS) KAKENHI Grant Nos.\ 18H01211 and 19K21874.
\end{acknowledgments}

\bibliographystyle{apsrev4-1}
\bibliography{spin}

\end{document}